\documentclass[prb,twocolumn,superscriptaddress,floatfix]{revtex4-2}
\usepackage{graphicx}
\usepackage{latexsym}
\usepackage{amsmath}
\usepackage{amsfonts}
\usepackage{amssymb}
\usepackage{multirow}
\usepackage{bm}
\usepackage{txfonts}
\usepackage{color}
\usepackage{braket}

\newcommand{\fig}[2]{\includegraphics[width=#1]{#2}}
\def\avs{{$A$V$_3$Sb$_5$}}

\def\bk{{\textbf{k}}}
\def\bM{{{\textbf{M}}}}
\def\br{{{\textbf{r}}}}

\def\bQ{{\textbf{Q}}}

\def\bG{{\textbf{G}}}
\def\ba{{{\bm a}}}
\newcommand{\nn}{{\langle ij \rangle}}

\usepackage{appendix}
\usepackage[colorlinks = true,
linkcolor = blue,
urlcolor  = blue,
citecolor = blue,
anchorcolor = blue]{hyperref}
\usepackage{ulem}
\usepackage{epstopdf}
\usepackage{multirow}

\begin{document}
\title{Interplay of Pair Density Wave and Uniform Superconductivity on the Kagome Lattice} 
\author{Ruiqing Fu}
\affiliation{Institute of Theoretical Physics, Chinese Academy of Sciences, Beijing 100190, China}
\affiliation{School of Physical Sciences, University of Chinese Academy of Sciences, Beijing 100049, China}

\author{Xianxin Wu}
\thanks{Contact Author: xxwu@itp.ac.cn}
\affiliation{Institute of Theoretical Physics, Chinese Academy of Sciences, Beijing 100190, China}

\author{Ziqiang Wang}
\thanks{Contact Author: wangzi@bc.edu}
\affiliation{Department of Physics, Boston College, Chestnut Hill, MA 02467, USA}

\author{Sen Zhou}
\thanks{Contact Author: zhousen@itp.ac.cn}
\affiliation{Institute of Theoretical Physics, Chinese Academy of Sciences, Beijing 100190, China}
\affiliation{School of Physical Sciences, University of Chinese Academy of Sciences, Beijing 100049, China}
	
\date{\today}
	
\begin{abstract}
We study the interplay between pair density wave (PDW) and uniform superconductivity on the kagome lattice within a single-orbital $t$-$W$-$J$ model near the $p$-type van Hove filling, where $W$ and $J$ denote effective pairing attractions on lattice sites and nearest-neighbor bonds, respectively. 
While onsite attraction favors zero-momentum pairing through the conventional BCS instability, nearest-neighbor attraction naturally promotes finite-momentum pairing due to the kagome geometry and sublattice interference. 
The model therefore provides a framework for exploring the competition and coexistence between uniform superconductivity and PDW order. 
Self-consistent mean-field calculations in a $2\times2$ supercell reveal that the uniform component reorganizes both the amplitudes and phase structure of the three $F_1$ PDW components, giving rise to superconducting states absent in the pure bond-pairing limit. 
A Ginzburg--Landau analysis identifies symmetry-allowed couplings that can favor this reorganization. 
The resulting states exhibit distinct charge modulations and quasiparticle spectra, while the chiral PDW undergoes topological transitions as the onsite attraction increases. 
Our results demonstrate how uniform superconductivity reshapes the order-parameter landscape and quasiparticle topology of a PDW state.
\end{abstract}
\maketitle

\section{Introduction}
The kagome metals \avs\ ($A$ = K, Rb, Cs) have recently emerged as an important platform for studying intertwined quantum orders. 
They host an unconventional charge density wave (CDW) order with signatures of time-reversal-symmetry (TRS) breaking and develop superconductivity at lower temperatures \cite{Ortiz2019, Wilson-PRL20, Yin2021, Mielke-Nat22, Jiang2023}. 
While the CDW state has been extensively investigated \cite{HuJP-SB21, Zhao2021CascadeCsV3Sb5, Leon-PRB21, LinYP-PRB21, HuJP-PRB21, Neupert-PRL21, Rafael-PRB22, Dong2023LoopCurrentCDW, SZ-NC22, Kontani-SA22, Jiang2023, WuXX-NSR25, WuXX-PRL26}, the nature of the superconducting phase remains under active debate. 
Experiments have reported fully gapped superconductivity with a variety of unconventional phenomena, including TRS-breaking, twofold anisotropy, higher-charge flux quantization, and signatures of pair density wave (PDW) \cite{Mu2021NQRCsV3Sb5, Cs2025PressureNematic, Zhong2023NodelessCsV3Sb5, Deng2024TRSBKagome, NbCs2025TRSB, Rb2023TwofoldSC, Fukushima2024ERS, Ge2024Charge46eCsV3Sb5, Chen2021RotonPDW, Zeljkovic-PRX23, Deng2024, Yan2024RbPDW, YinJX-PRB26, YinJx-PNAS2026}.
In particular, scanning tunneling microscopy (STM) and Josephson STM measurements have revealed spatially modulated superconducting responses consistent with finite-momentum pairing \cite{Chen2021RotonPDW, Zeljkovic-PRX23, Deng2024, Yan2024RbPDW, YinJX-PRB26, YinJx-PNAS2026}, suggesting that PDW physics may play an important role in kagome superconductors.

PDW has been extensively studied in the context of cuprate superconductors, and been subsequently extended to a broad range of correlated materials, including iron-based superconductors, transition-metal dichalcogenides, heavy-fermion systems, and kagome superconductors \cite{Himeda-PRL02, ZhangSC-PRL07, Lee-PRX14, Dai2020QuantumDisorderedPDW, Agterberg2020, Hamidian2016CupratePDW, Ruan2018CupratePDW, Edkins2019VortexHaloPDW, Choubey2020CupratePDWDSC, Liu2023FeSePDW, Zhao2023SmecticPDW, Kong2025PairDensityModulation, Liu2021NbSe2CooperPDW, Gu2023UTe2PDW, Chen2021RotonPDW, LeeSB-SPC23, Deng2024, Yan2024RbPDW, GaoHJ-FP26}. 
It is an unconventional superconducting state in which Cooper pairs condense at finite center-of-mass momenta, leading to a spatially varying superconducting order parameter even in the absence of an external magnetic field \cite{Agterberg2008DislocationsPDW, Agterberg2020, GaoHJ-FP26, YinJX-NRP26}. 
Because PDW simultaneously breaks gauge and translational symmetries, it naturally intertwines superconductivity with a variety of secondary orders and collective phenomena, including charge and spin modulations, nematicity, loop-current order, higher-charge superconductivity, and residual low-energy quasiparticles \cite{Berg-PRB09, Berg2009Charge4e, Agterberg-PRB11, Agterberg2015LoopCurrent, Dai2020QuantumDisorderedPDW, Jin2022NestedFSPDW, SZ-NC22, Wu2024Charge4ePDW}.
Therefore, PDW provides a unified framework for understanding unconventional superconducting states with intertwined broken symmetries.

From a microscopic perspective, finite-momentum pairing is generally disfavored in the weak-coupling limit relative to zero-momentum pairing, which benefits from the logarithmic Cooper instability.
Although it has been shown that PDW can be promoted by nested pairing channels, multiband form factors, and lattice-specific electronic structures \cite{Berg-PRB09, Lee-PRX14, YaoH-PRL20, Jin2022NestedFSPDW, Yafis-PRL23, Raghu-PRB23, Ticea2024MultibandPDW, YaoH-PRL24, Kee-arXiv26, WangQH-arXiv26}, the spontaneous emergence of PDW is rare in microscopic models.
Interestingly, because of the unique geometry and sublattice quantum interference effect of the kagome lattice, the susceptibilities of pairings on nearest-neighbor (NN) bonds are significantly enhanced at the three inequivalent van Hove (vH) points, near the $p$-type vH filling.
Consequently, attractive pairing interactions on NN bonds can naturally lead to triple-\bQ\ $2\times2$ bond PDW \cite{Yao2024PDW}. 
Depending on the relative phases of the three ordering components, this state can realize either real or chiral configurations. 
However, existing microscopic studies have primarily focused on the formation of pure PDW states \cite{SZ-NC22, Mohanta-PRB23, Yao2024PDW, WangQH-arXiv26}, while experimentally, the PDW modulation is shown to coexist with uniform superconductivity \cite{Chen2021RotonPDW, Deng2024, Yan2024RbPDW, YinJx-PNAS2026}. 
Therefore, it is natural to ask how finite-\bQ\ PDW and zero-\bQ\ superconductivity compete and coexist.
In particular, understanding how a uniform superconducting component modifies the PDW phase diagram, quasiparticle spectrum, topology, and associated secondary orders remains an important open problem.

In this work, we study an effective pairing model with attractive interactions on both lattice sites and NN bonds on the kagome lattice near the $p$-type vH filling.
The onsite and NN pairing attractions naturally favor onsite uniform superconductivity and $2\times2$ bond PDW, respectively, thereby enabling us to investigate the competition and coexistence of PDW and uniform superconductivity on the kagome lattice.
Combining self-consistent mean-field calculations and extended point-group symmetry analysis, we demonstrate that the interplay between finite- and zero-momentum pairing qualitatively reshapes the PDW landscape.
Besides the previously identified real and chiral PDW states, we find two coexistence-induced phases: a nematic PDW and a phase-twisted PDW.
A Ginzburg-Landau analysis identifies symmetry-allowed couplings that help explain their order-parameter structures.
We further show that, along a representative parameter cut, increasing onsite attraction drives successive Chern-number transitions within the chiral PDW.
The phases also exhibit characteristic low-energy spectral signatures.

\begin{figure}
\begin{center}
\fig{3.4in}{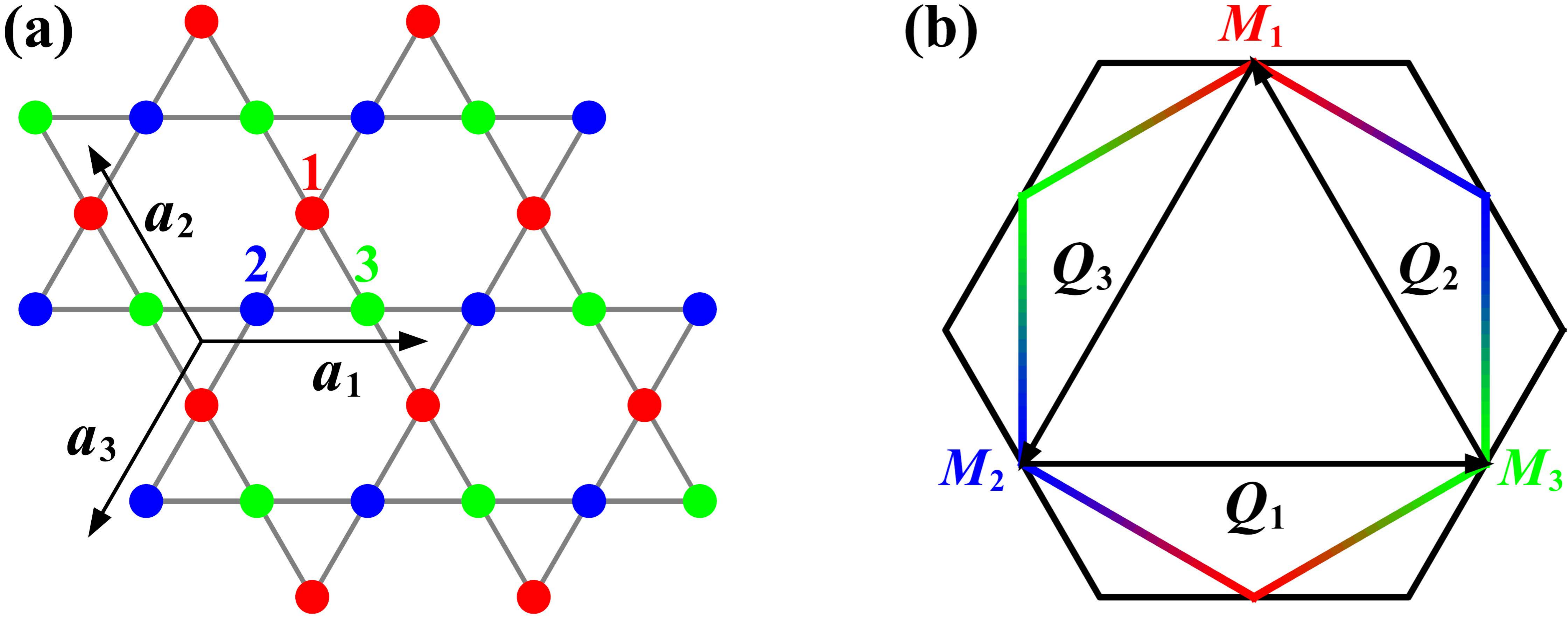} 
\caption{(a) Kagome lattice with three sublattices denoted by red (1), blue (2), and green (3) circles. The two basis lattice vectors are $\ba_1 = (1, 0)$ and $\ba_2 = (-\tfrac{1}{2}, \tfrac{\sqrt{3}}{2})$, with the third direction defined as $\ba_3 =-\ba_1 -\ba_2$. (b) Sublattice-resolved FS at $p$-type vH filling in the hexagonal BZ. Colors of lines display the sublattice contents. The three independent vH points are labeled by $\bM_{1,2,3} = \tfrac{1}{2} \bG_{1,2,3}$, with the reciprocal lattice vectors given by $\bG_1 = (0, \tfrac{4\pi}{\sqrt{3}})$, $\bG_2 = (-2\pi, -\tfrac{2\pi}{\sqrt{3}})$, and $\bG_3 = (2\pi, -\tfrac{2\pi}{\sqrt{3}})$.}  \label{fig1} 
\end{center}
\vskip-0.8cm
\end{figure}

\section{Model and Method}
We consider the effective single-orbital $t$-$W$-$J$ model on the kagome lattice depicted in Fig. \ref{fig1}(a),
\begin{equation}
\hat{H} = -t\sum_{\nn,\sigma} (c^\dagger_{i\sigma} c_{j\sigma} +\text{H.c.}) -W\sum_i  \hat{\Delta}_i^\dagger \hat{\Delta}_i -J\sum_{\nn} \hat{\Delta}_{ij}^\dagger \hat{\Delta}_{ij}, \label{Ham}
\end{equation}
where $c^\dagger_{i\sigma}$ creates a spin-$\sigma$ ($\uparrow$, $\downarrow$) electron at site $i$, and $t$ denotes the NN hopping which is set as the energy unit, $t=1$.
The interactions $J$ and $W$, both positive, describe the phenomenological attractions for spin-singlet pairings on NN bonds $\hat{\Delta}_{ij} = c_{i\uparrow} c_{j\downarrow} -c_{i\downarrow} c_{j\uparrow}$ and lattice sites $\hat{\Delta}_i = \hat{\Delta}_{ii}=2c_{i\uparrow}c_{i\downarrow}$, respectively.
Decoupling the pairing interactions in terms of $\hat{\Delta}_i$ and $\hat{\Delta}_{ij}$, one reaches the mean-field Hamiltonian,
\begin{align}
\hat{H}_\text{MF} = & -t\sum_{\nn,\sigma} (c^\dagger_{i\sigma} c_{j\sigma} +\text{H.c.}) 
-W\sum_i \left( \Delta^*_i \hat{\Delta}_i +\text{H.c.} - |\Delta_i|^2 \right) \nonumber \\
& -J\sum_{\nn} \left( \Delta^*_{ij} \hat{\Delta}_{ij} +\text{H.c.} - |\Delta_{ij}|^2 \right). \label{Hmf}
\end{align}
The order parameters, the onsite pairing $\Delta_i = \langle \hat{\Delta}_i \rangle$ and the NN pairing $\Delta_{ij} =\langle \hat{\Delta}_{ij} \rangle$, are to be determined self-consistently by solving the corresponding Bogoliubov-de Gennes equations.
We consider the quantum states to be periodic with an enlarged $2\times 2$ unit cell, so as to accommodate both the uniform superconductivity driven by onsite $W$ and the PDW phase promoted by NN pairing $J$.
To explore competing states within the $2\times 2$ variational space, we use different initial conditions for solving the self-consistent equations numerically.
When multiple states converge at a given set of interactions, we compare their energies to identify the lowest-energy state among those searched.

To characterize the superconducting states that are periodic in a $2\times 2$ enlarged unit cell, which contains 12 sites and 24 NN bonds, we analyze the order parameters within the framework of the extended point group $C'''_{6v}$, which is the natural symmetry group for ordering phenomena associated with the three inequivalent vH points ($\bM_\alpha$, $\alpha$ = 1, 2, 3) on the kagome lattice \cite{Venderbos2016ExtendedPointGroup, Wagner2023BondFluxKagome, PhysRevResearch.7.023129}.
Unlike the ordinary point group $C_{6v}$, the extended point group incorporates translations that become symmetry operations in the enlarged $2\times2$ unit cell and therefore provides a unified classification of translation-symmetry-preserving and translation-symmetry-breaking orders.
As a consequence, in addition to the four one-dimensional \{$A_1$, $A_2$, $B_1$, $B_2$\} and two two-dimensional \{$E_1$, $E_2$\} irreducible representations inherited from $C_{6v}$, the extended point group $C'''_{6v}$ contains four three-dimensional irreducible representations $F_m$, $m=1,\cdots,4$, which explicitly break the translation symmetry of the pristine kagome lattice and describe orderings at the wave vectors $\bQ_\alpha=\bM_\alpha$.
The self-consistently obtained pairing order parameters, onsite $\Delta_i$ and NN bond $\Delta_{\langle ij\rangle}$, are then projected separately onto normalized basis patterns of $C'''_{6v}$ \cite{PhysRevResearch.7.023129}.
The decomposition in each irreducible representation represents the superconducting order parameter in the corresponding pairing channel.

We find that the dominant channels for both onsite and NN pairings are the one-dimensional $A_1$, two-dimensional $E_2$, and three-dimensional $F_1$ representations, in the representative low-energy states we explored. 
Smaller components remain in the full self-consistent fields are retained in the calculating physical observables.
The decomposition in $A_1$, $E_2$, and $F_1$ irreducible representations provides the pairing order parameters in the corresponding channels and are denoted as, respectively, $\psi^p$, ${\bm \xi}^p$ = $(\xi^p_1, \xi^p_2)$, and ${\bm \eta}^p$ = $(\eta^p_1, \eta^p_2, \eta^p_3)$, where $p=s$ for onsite pairing and $p=b$ for NN bond pairing.
Consequently, the onsite pairing on sublattice $\alpha$ (= 1, 2, 3) in the pristine unit cell at $\br$, \textit{i.e.}, site $i=(\br, \alpha)$, is given by
\begin{equation}
\Delta^s_{\br\alpha} \simeq \psi^s +\xi^s_1 e_{1,\alpha}+\xi^s_2 e_{2,\alpha}  +\sqrt{3} \eta^s_\alpha \cos(\bQ_\alpha\cdot \br), \label{Dsite}
\end{equation}
where the $E_2$ basis vectors, normalized over the sites or bonds of the supercell, are
\begin{equation}
{\bm e}_1=\tfrac{1}{\sqrt{2}}(-1,-1,2),\quad {\bm e}_2=\sqrt{\tfrac{3}{2}}(-1,1,0).
\end{equation}
Along each $\ba_\alpha$ direction, there are two independent NN bonds, denoted by $\nu=\pm 1$, within a pristine unit cell.
The $\nu=+1$ NN bonds connect lattice sites ($\br$, $\beta$) and ($\br$, $\gamma$), with indices $(\alpha,\beta,\gamma)$ = (1, 2, 3), (2, 3, 1), or (3, 1, 2). 
They form the up-triangles of the kagome lattice.
On the other hand, the $\nu=-1$ NN bonds connect lattice sites ($\br$, $\beta$) and ($\br-\ba_\alpha$, $\gamma$) and produce the down-triangles in the kagome lattice.
The pairing on $\nu$ NN bond along $\ba_\alpha$ direction in the pristine unit cell at $\br$ is then given by
\begin{equation}
\Delta^b_{\br\alpha\nu} \simeq \psi^b+\xi^b_1 e_{1,\alpha}+\xi^b_2 e_{2,\alpha}+\nu \sqrt{3} \eta^b_\alpha \cos(\bQ_\alpha \cdot \br). \label{Dbond}
\end{equation}
We note that the NN bond space contains a second, one to two order of magnitude smaller copy of $F_1$ \cite{PhysRevResearch.7.023129}; it is not included in Eq.~\eqref{Dbond} but is retained in the full numerical fields.
The pairing order parameters $\psi^{s/b}$, $\xi^{s/b}_\alpha$, and $\eta^{s/b}_\alpha$ are in general complex, encoding the relative phases among different pairing components.
We note that superconducting state has a global $U(1)$ phase, which we fix by choosing $\psi^s$ to be real and positive in the rest of the paper.
This symmetry decomposition provides a compact description of the self-consistent solutions and forms the basis for identifying various PDW phases discussed below. 
The lattice conventions, exact projections, and residual components are given in Appendix~\ref{app:symmetry}.

\begin{figure}
\begin{center}
\fig{3.4in}{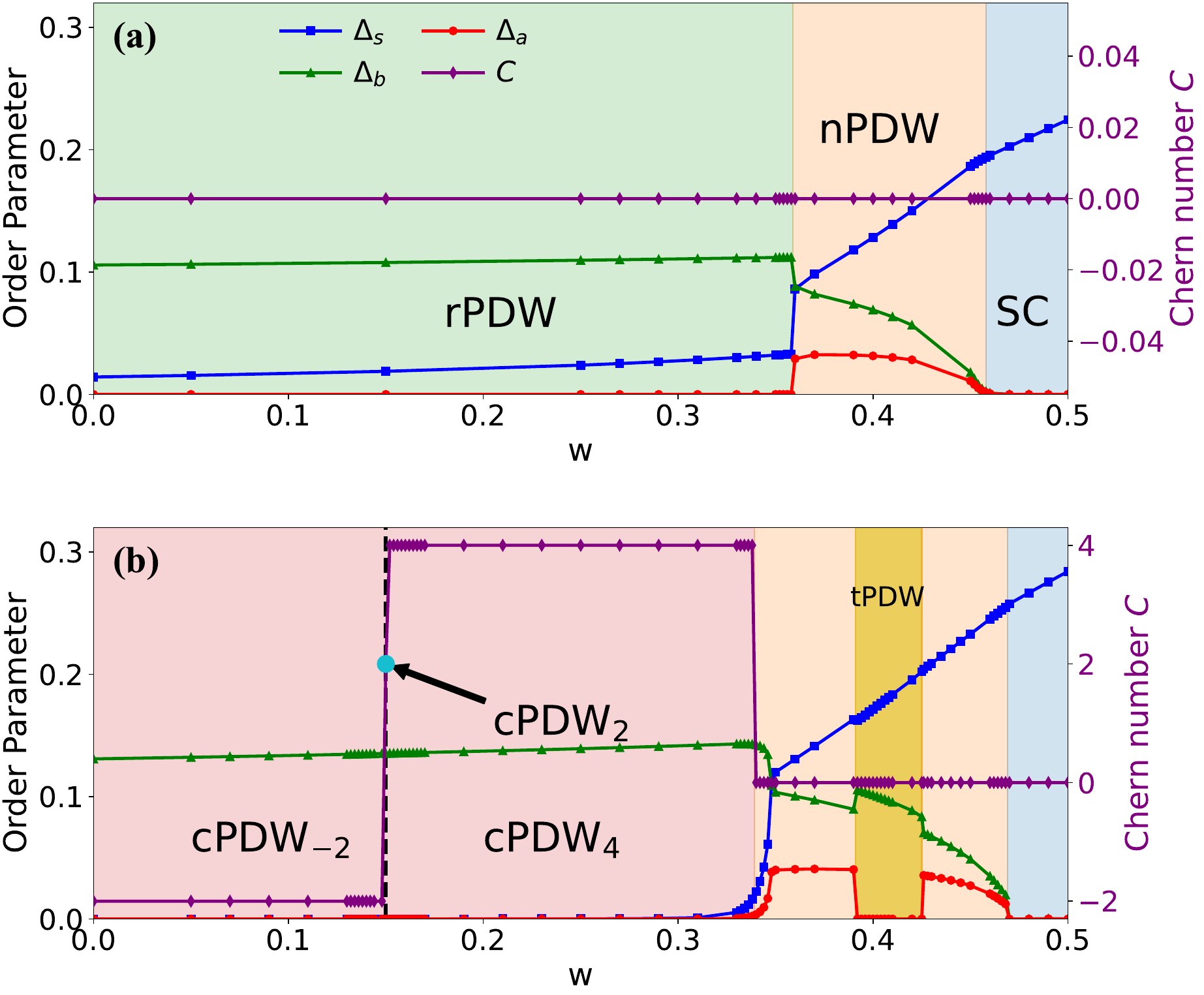} 
\caption{State evolution and its Chern number as a function of onsite attraction $W$ along two fixed-$(n,J)$ cuts: (a) $(n,J)$ = (5/12, 1.3) and (b) (5.7/12, 1.5). Shading identifies the rPDW, cPDW, nPDW, and tPDW states and the uniform SC. The cPDW subscript gives $C$; the cyan marker in (b) highlights the narrow cPDW$_2$ interval.} \label{fig2}
\end{center}
\vskip-0.8cm
\end{figure}

\section{Results}
\subsection{Phase Diagrams and Phase Transitions}
Before discussing the interplay between PDW and uniform superconductivity, we first briefly review the properties of the pure bond-PDW obtained in the absence of on-site pairing attraction, $W$ = 0.
At $p$-type vH filling, $n=5/12$, the Fermi surface of noninteracting electrons is the hexagon that connects the three inequivalent vH points $\bM_\alpha$ at the Brillouin-zone boundary, as shown in Fig. \ref{fig1}(b). 
Any two vH points $\bM_\alpha$ and $\bM_\beta$ are nested by the third $2\times2$ ordering vector $\bQ_\gamma = \bG_\gamma/2$, with the indices obeying cyclic permutation $\epsilon_{\alpha\beta\gamma}=1$. 
More importantly, the kagome lattice is special in that the Bloch states at the vH points $\bM_\alpha =\bQ_\alpha$ are exclusively localized on the $\alpha$th sublattice due to the sublattice interference effect \cite{LiJX-PRB12, WangQH-PRB13, WuXX-PRL21, SZ-NC22, Dong2023LoopCurrentCDW}, as illustrated in the sublattice resolved Fermi surface in Fig. \ref{fig1}(b). 
Consequently, pairing on an NN bond connecting sublattices $\alpha$ and $\beta$ naturally couples electronic states associated with different vH points and carries the finite center-of-mass momentum $\bQ_\gamma$. 
This microscopic structure strongly promotes $2\times2$ bond PDW order near the vH filling \cite{Yao2024PDW}.

In the pure bond-pairing problem, the three symmetry-related PDW components have equal amplitudes, $|\eta^b_\alpha|=\eta$, and their relative phases distinguish two parent triple-\bQ\ PDW states. 
The real PDW (rPDW) state, ${\bm\eta}^b =\eta (1,1,1)$, preserves TRS and sixfold rotation, whereas the chiral PDW (cPDW) state, ${\bm\eta}^b =\eta (1, e^{i2\pi/3}, e^{i4\pi/3})$, has a dominant $F_1$ triplet that preserves rotation up to a gauge phase but breaks TRS through its $2\pi/3$ phase winding. 
Their competition in the $n$-$J$ plane has been established in Ref.~\cite{Yao2024PDW}. We therefore choose two representative cuts that start from different parent PDW states: $(n,J)=(5/12,1.3)$ in the rPDW regime and $(5.7/12,1.5)$ in the cPDW regime. By increasing $W$ along these cuts, we can directly follow how a growing tendency toward onsite uniform pairing competes and coexists with the bond PDW.

\begin{figure*}
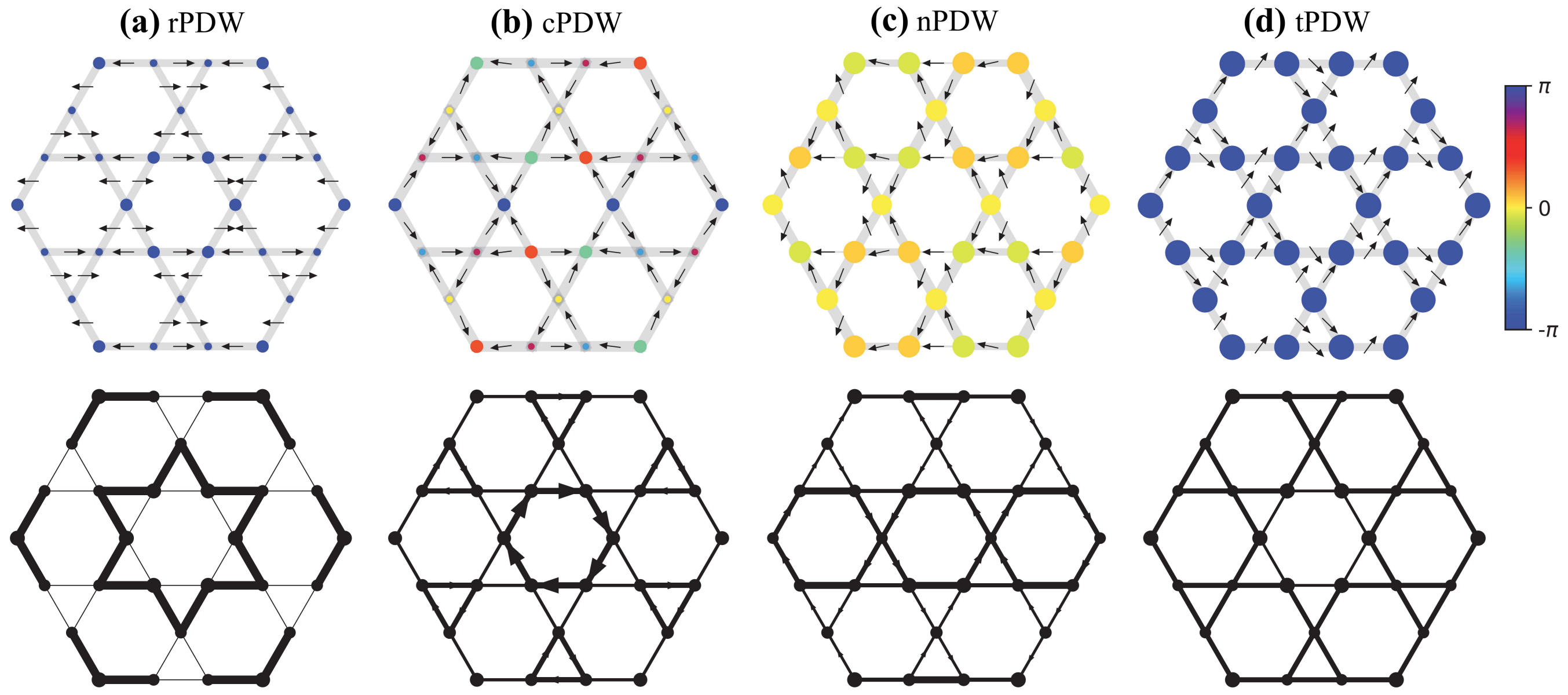

\begin{center}
\fig{7in}{fig3.pdf} 
\caption{Representative real-space pairing texture (upper panels) and induced charge and bond responses (lower panels) of the four PDW phases: (a) rPDW, (b) cPDW, (c) nPDW, and (d) phase-twisted PDW (tPDW). 
In upper panels, the size and the color of the dots denote the magnitude and phase of the on-site pairing order $\Delta_i$, while the bond thickness and the arrow direction represent the magnitude and phase of the bond pairing order $\Delta_{ij}$. In the lower panels, dot size and bond thickness represent onsite density and bond correlations, while arrows represent bond currents. 
The data are taken from representative self-consistent solutions at ($n$, $J$, $W$) = (5/12,1.3,0.25) for rPDW, (5.7/12, 1.5, 0.19) for cPDW, (5.7/12, 1.5, 0.36) for nPDW, and (5.7/12, 1.5, 0.41) for tPDW.} \label{fig3}
\end{center}
\vskip-0.8cm
\end{figure*}

The resulting evolution is summarized in Fig.~\ref{fig2}. 
The pairing texture of onsite $\Delta_i$ and NN bond $\Delta_{\langle ij \rangle}$, as well as their decomposition in the extended point group $C'''_{6v}$, will be discussed in Subsection \ref{sec3b}.
Here, we focus on four quantities that help us characterize the resulting phases and identify the corresponding phase transitions: the uniform onsite $A_1$ amplitude $\Delta_s=|\psi^s|$, the total amplitude of the bond $F_1$ PDW order $\Delta_b=\sum_\alpha|\eta^b_\alpha|$, its amplitude anisotropy $\Delta_a=\sum_\alpha\big||\eta^b_\alpha|- \tfrac{1}{3}\Delta_b \big|$, and the Chern number $C$ of the occupied Bogoliubov bands in gapped states. 
At small $W$, the bond PDW remains dominant and the parent states survive essentially unchanged: rPDW persists to $W\simeq0.36$ for $(n,J)=(5/12,1.3)$, while cPDW persists to $W\simeq0.35$ for $(5.7/12,1.5)$. 
The cPDW regime is further divided into topologically distinct sectors with $C=-2$, $2$, and $4$. 
These changes of $C$ occur within the same cPDW branch and will be traced to the closing and reopening of quasiparticle gaps in Subsection \ref{sec3c}.

The most important feature of Fig.~\ref{fig2} is the intermediate regime in which the onsite and bond pairing channels become comparable. 
Rather than evolving directly from a parent PDW to uniform superconductivity, the system develops additional minima of the superconducting free energy. 
On both parameter cuts, the equal-amplitude triple-\bQ\ state first gives way to a nematic PDW (nPDW), signaled by a finite $\Delta_a$: one PDW component acquires an amplitude different from the other two and the sixfold rotational symmetry is spontaneously broken.
On the cut starting from cPDW in Fig.~\ref{fig2}(b), a narrow phase-twisted PDW (tPDW) region also appears between two nPDW regions. 
In tPDW, the three bond-PDW components recover equal amplitudes and are mutually in phase, but acquire a nontrivial phase relative to the uniform onsite condensate. 
Thus nPDW and tPDW are not continuations of the pure bond-PDW solutions; they are generated specifically by the coexistence and competition of finite- and zero-momentum pairing.

At still larger $W$, $\Delta_s$ grows rapidly while $\Delta_b$ is suppressed, and the PDW eventually disappears at $W\simeq0.46$ and $0.48$ for the two cuts, respectively. The resulting state is a translation-invariant $A_1$ superconductor with uniform onsite and NN bond pairings. 
The evolution near this boundary is comparatively smooth: $\Delta_s$ changes continuously while $\Delta_b$ decreases to zero, consistent with a continuous or weakly first-order transition within our numerical resolution. 
By contrast, transitions among distinct PDW states are accompanied by discontinuous changes of the order parameters in Fig.~\ref{fig2}, indicating first-order transitions. 
The two cuts therefore reveal a common physical progression: a PDW-dominated regime at small $W$, an intermediate coexistence regime in which new symmetry-broken states emerge, and a uniform-superconducting regime at large $W$. 
Next, we examine the structures of pairing order parameters and the corresponding charge responses that distinguish these phases.

\begin{table*}[htb]
	\caption{Order-parameter components of the four representative PDW phases in Fig.~\ref{fig3}, resolved in the analytic basis of the $A_1$, $E_2$, and $F_1$ representations using the lattice convention of the main calculation. The basis functions are normalized as $\sum_l |f_l|^2=N$, with $N=12$ for onsite pairing and $N=24$ for NN bond pairing. The coefficients are obtained by the orthogonal projection $\langle f,\Delta\rangle/N$. The global $U(1)$ phase is chosen such that the onsite $A_1$ component is real and positive; for cPDW the tiny onsite $A_1$ component is numerically phase aligned with $\eta^s_1$. The bond $F_1$ columns contain its dominant copy. The coefficients are anomalous averages, not pairing potentials; the first and second rows for each state give their magnitudes and phases, respectively.}\label{tab1}
	\begin{ruledtabular}
		\begin{tabular}{c|ccc|ccc}
			\multirow{2}{*}{State}
			& \multicolumn{3}{c|}{Onsite pairing $\Delta_i$}
			& \multicolumn{3}{c}{NN bond pairing $\Delta_{ij}$} \\
			& $A_1$ ($\psi^s$) & $E_2$ (${\bm \xi}^s$) & $F_1$ (${\bm \eta}^s$)
			& $A_1$ ($\psi^b$) & $E_2$ (${\bm \xi}^b$) & $F_1$ (${\bm \eta}^b$) \\
			\hline
			
				\multirow{2}{*}{rPDW}
				& 0.0239 & 0 & 0.00747 $\times$ (1, 1, 1)
				& $1.98{\times}10^{-4}$ & 0 & 0.0366 $\times$ (1, 1, 1) \\
				& 0 &  & ($\pi$, $\pi$, $\pi$)
				& $\pi$ &  & ($\pi$, $\pi$, $\pi$) \\
			\hline
			
				\multirow{2}{*}{cPDW}
				& $5.95{\times}10^{-5}$
				& 0.0118 $\times$ (1, 1)
				& 0.0143 $\times$ (1, 1, 1)
				& $1.00{\times}10^{-5}$
				& 0.00560 $\times$ (1, 1)
				& 0.0456 $\times$ (1, 1, 1) \\
					& 0
					& $(\pi/3,\,-\pi/6)$
					& $(0,\,2\pi/3,\,4\pi/3)$
					& $\pi$
					& $(\pi/3,\,-\pi/6)$
					& $(0,\,2\pi/3,\,4\pi/3)$ \\
			\hline
			
				\multirow{2}{*}{nPDW}
				& 0.131
				& (0.00299, 0.00518)
				& (0.00440, 0.0163, 0.0163)
				& 0.0318
				& (0.00120, 0.00208)
				& (0.0131, 0.0436, 0.0436) \\
				& 0
				& (0, 0)
				& (0, $\pi/2$, $-\pi/2$)
				& $\pi$
				& (0, 0)
				& (0, $\pi/2$, $-\pi/2$) \\
			\hline
			
				\multirow{2}{*}{tPDW}
				& 0.184 & 0 & 0.0123 $\times$ (1, 1, 1)
				& 0.0434 & 0 & 0.0318 $\times$ (1, 1, 1) \\
				& 0 &  & 0.546$\pi \times$ (1, 1, 1)
				& $\pi$ &  & 0.559$\pi \times$ (1, 1, 1) \\
		\end{tabular}
	\end{ruledtabular}
\end{table*}

\subsection{Pairing Structures and Charge Modulations} \label{sec3b}
To expose the physical differences among the four PDW phases identified in Fig. \ref{fig2}, we analyze both their real-space textures and their symmetry-resolved order parameters, and compute the corresponding charge responses.
The upper panels of Fig.~\ref{fig3} show representative onsite and NN-bond pairing configurations, while the lower panels display the associated charge and bond modulations.
In the upper panels, the dot size and color denote the amplitude and phase of on-site pairing $\Delta_i$, while the bond thickness and the arrow direction represent the amplitude and phase of NN bond pairing $\Delta_{\langle ij\rangle}$.
In the lower panels, the dot size represents the onsite density $n_i=\sum_\sigma\langle c^\dagger_{i\sigma}c_{i\sigma}\rangle$, the line thickness represents Re$\chi_{\langle ij \rangle}$ with $\chi_{\langle ij \rangle}=\sum_\sigma\langle c^\dagger_{i\sigma} c_{j\sigma} \rangle$, and the arrows represent bond currents proportional to $\mathrm{Im} \chi_{\langle ij \rangle}$. 
The four examples are chosen from the corresponding regions of Fig.~\ref{fig2}: $(n,J,W)$ = (5/12, 1.3, 0.25) for rPDW in Fig. \ref{fig3}(a), (5.7/12, 1.5, 0.19) for cPDW in Fig. \ref{fig3}(b), (5.7/12, 1.5, 0.36) for nPDW in Fig. \ref{fig3}(c), and (5.7/12, 1.5, 0.41) for tPDW in Fig. \ref{fig3}(d), respectively.

A compact characterization is obtained by projecting the self-consistent pair fields onto the $A_1$, $E_2$, and $F_1$ representations of $C'''_{6v}$, as summarized in Table~\ref{tab1}. 
In all four states the largest bond component is the three-component $F_1$ order ${\bm\eta}^b$, confirming that they remain triple-\bQ\ PDW states even when the uniform channel becomes sizable. 
Their coexistence, however, induces additional onsite and bond components whose symmetry is constrained by the primary PDW configuration. 
In particular, commensurability permits symmetry- and gauge-allowed couplings between PDW composites and leads to zero-momentum superconducting components \cite{SZ-NC22, ZhangGM-PRB25}. 
This is already visible in the parent phases: rPDW induces an onsite $A_1$ component, whereas the leading uniform onsite component of cPDW has $E_2$ symmetry. 
Consequently, as shown in Fig. \ref{fig2}, the uniform onsite $A_1$ amplitude $\Delta_s=|\psi^s|$ is finite in rPDW even at $W=0$ and increases with $W$, while it remains extremely small in cPDW; for example, $\Delta_s \simeq 1.4\times10^{-4}$ at $W$ = 0.25 on the cPDW cut.

For rPDW, the dominant bond $F_1$ order is proportional to $(1,1,1)$, so the three ordering vectors enter with equal amplitudes and the same phase. 
The real-space pattern in Fig.~\ref{fig3}(a) therefore retains both TRS and the sixfold rotational symmetry. 
The accompanying density and bond modulations inherit the same rotational symmetry. 
The cPDW state has an equally symmetric amplitude structure but a qualitatively different phase configuration, ${\bm\eta}^b\propto(1,e^{i2\pi/3},e^{i4\pi/3})$.
The phase winding breaks TRS and produces finite circulating bond currents, as directly seen in the lower panel of Fig.~\ref{fig3}(b), while the equal amplitudes preserve the rotational character of the dominant $F_1$ order. 
The full cPDW state also contains uniform $E_2$ components and a much smaller $A_1$ component (Table~\ref{tab1}). 
Since these components transform with different phases under rotation, their coexistence permits weak rotational-symmetry breaking in cPDW even while the dominant $F_1$ amplitudes remain equal. 
Thus the distinction between the two parent phases resides primarily in the internal phase structure of the triple-\bQ\ order rather than in its spatial amplitude pattern.

The coexistence-induced nPDW reorganizes both the amplitudes and phases of the PDW components. 
For the representative solution in Table~\ref{tab1}, the onsite and bond $F_1$ component ${\bm \eta}$ = $(a, ib, -ib)$ with $a,b$ positive real numbers, which selects one crystalline direction and produces the pronounced nematic pairing and charge textures showing in Fig.~\ref{fig3}(c). 
Under complex conjugation its $F_1$ components transform as $(\eta_1, \eta_2, \eta_3) \rightarrow (\eta_1, -\eta_2, -\eta_3)$, and these sign changes can be compensated by translation along ${\bm a}_1$. 
Therefore, the nPDW breaks ordinary time-reversal but preserves time-reversal followed by a primitive translation, which is consistent with its vanishing Chern number.
The unequal PDW amplitudes break sixfold rotation in nPDW.

The tPDW state realizes the complementary possibility. 
Its bond $F_1$ components again have equal amplitudes and are mutually in phase, restoring the rotational character of the triple-\bQ\ PDW phase. 
However, there is a nontrivial phase twist between the bond $F_1$ and the uniform onsite $A_1$ condensate, which is $0.559\pi$ in the representative solution listed in Table~\ref{tab1}. 
Because this relative phase cannot be removed by a global gauge transformation, it breaks TRS even though the PDW sector itself has no $2\pi/3$ chiral winding. 
The tPDW is therefore physically distinct from cPDW: cPDW breaks TRS through relative phases among the three finite-\bQ\ components, whereas tPDW breaks TRS through the phase relation between the finite-\bQ\ PDW and zero-momentum condensates.
This difference in phase structure may underlie their distinct topological behavior: cPDW has a nonzero Chern number in its gapped sectors, whereas tPDW remains at $C=0$ throughout the region found here despite breaking TRS.

The symmetry analysis thus reveals the organizing principle behind the intermediate phases: coexistence does not simply superimpose uniform superconductivity on an unchanged PDW. 
In nPDW, the coupling reorganizes the amplitudes and relative phases of the three PDW components, producing a nematic state. 
In tPDW, the PDW components remain equal and mutually in phase, but their phase relative to the uniform condensate breaks TRS. 
These distinct order-parameter structures are imprinted on the induced charge and current patterns in Fig.~\ref{fig3} and, as we show next, on the low-energy quasiparticle spectra.

\begin{figure*}
\begin{center}
\fig{7in}{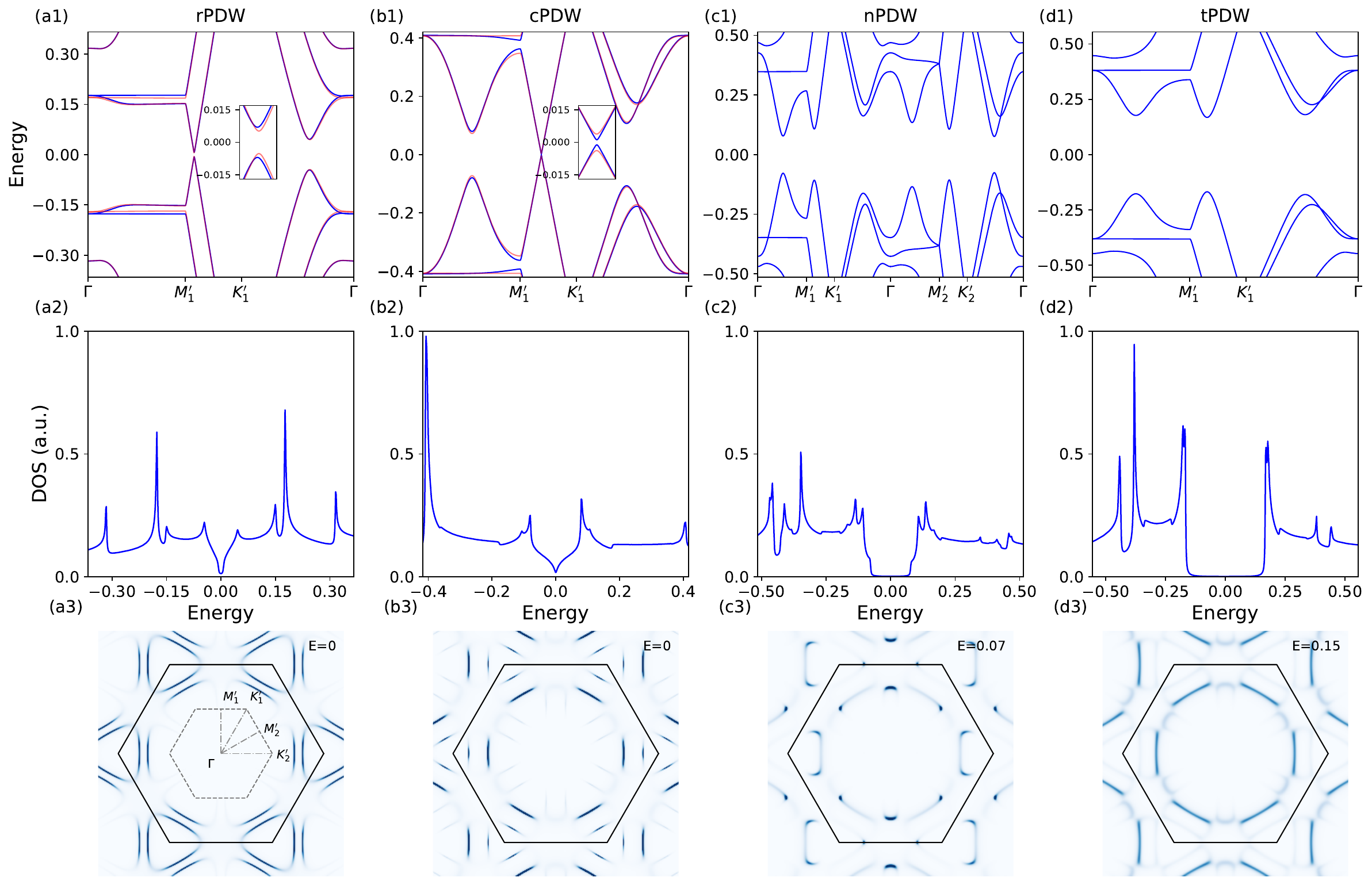} 
\caption{Representative quasiparticle properties of the four PDW phases: (a) rPDW at $(n,J,W)$ = (5/12, 1.3, 0.25), (b) cPDW at (5.7/12, 1.5, 0.19), (c) nPDW at (5.7/12, 1.5, 0.36), and tPDW at (5.7/12, 1.5, 0.41). 
(a1)-(d1) The Bogoliubov band structures along the high-symmetry path in the $2\times2$ reduced BZ. In (a1) and (b1), the faint red line denotes the band structure of cPDW and rPDW at the same parameters, and the inset zooms into the small gaps near the Bogoliubov-Dirac crossing. 
(a2)-(d2) The corresponding DOS. 
(a3)-(d3) The constant-energy spectral intensities at $0$, $0$, $0.07t$, and $0.15t$, respectively. 
The density of states and spectral intensities are shown in arbitrary units, and a thermal broadening of $k_BT = 0.002t$ is used in their computations.} \label{fig4}
\end{center}
\vskip-0.8cm
\end{figure*}

\subsection{Quasiparticle Properties and Topology} \label{sec3c}
The different order-parameter structures discussed above lead to characteristic reconstructions of the Bogoliubov quasiparticles. 
Fig. \ref{fig4} compares the low-energy band dispersions, density of states (DOS), and constant-energy spectral intensities of the four representative PDW states in Fig. \ref{fig3}. 
The rPDW and cPDW states provide a useful comparison: their translation-invariant pairing components are relatively small and their dominant bond $F_1$ amplitudes are similar, whereas their internal phases differ fundamentally. 
Their overall dispersions are nevertheless remarkably close [Figs. \ref{fig4}(a1) and \ref{fig4}(b1)], as emphasized by the overlaid metastable solutions at the same parameter points. 
In both states, the $2\times2$ triple-\bQ\ $F_1$ components open substantial gaps on the nested portions of the normal-state Fermi surface, while much smaller gaps remain near the Bogoliubov-Dirac crossings.
These small gaps leave appreciable low-energy weight in the DOS [Figs. \ref{fig4}(a2) and \ref{fig4}(b2)] and, with spectral broadening, arc-like zero-energy intensity confined to restricted momentum regions [Figs. \ref{fig4}(a3) and \ref{fig4}(b3)]. 
The residual arcs qualitatively resemble those observed in kagome superconductors \cite{Deng2024}, although broadened zero-energy intensity does not by itself imply an exactly gapless spectrum.

The quasiparticle spectra change visibly once the uniform component becomes comparable to the PDW. 
In the coexistence phases nPDW and tPDW, the states near the Bogoliubov-Dirac crossings of the parent PDWs are more strongly gapped [Figs. \ref{fig4}(c1) and \ref{fig4}(d1)], and the corresponding DOS is U-shaped, with strongly suppressed low-energy weight [Figs. \ref{fig4}(c2) and \ref{fig4}(d2)]. 
In nPDW, unequal PDW amplitudes remove the equivalence among the three crystalline directions, producing direction-dependent bands and anisotropic spectral intensity outside the gap [Figs. \ref{fig4}(c1) and \ref{fig4}(c3)]. 
This momentum-space anisotropy provides a direct spectroscopic manifestation of the nematic order identified from the real-space texture. 
In tPDW, by contrast, the three PDW amplitudes remain symmetry related, and the finite-energy spectral map retains the rotational character of the dominant PDW order [Fig.~\ref{fig4}(d3)].

These spectral features provide experimentally accessible signatures for distinguishing the PDW phases. 
The PDW-dominated rPDW and cPDW states exhibit gapped Bogoliubov-Dirac crossings, a roughly V-shaped DOS outside their small gaps, and arc-like remnants of the reconstructed Fermi surface. 
The coexistence nPDW and tPDW states instead show larger low-energy gaps and a U-shaped DOS, while the momentum-space anisotropy of nPDW reveals its broken rotational symmetry. 
These two coexistence phases may be further distinguished by their real space gap modulation: the unequal PDW amplitudes in nPDW lead to anisotropic modulation strengths along the three crystalline directions, whereas the symmetry-related components in tPDW preserve the rotational symmetry of the gap modulations.
The quasiparticle spectra of rPDW and cPDW remain remarkably similar, even near their small Bogoliubov-Dirac gaps. 
Their distinct internal phase structures are instead reflected in the Berry curvature and topology of the Bogoliubov bands, as we discuss next.

\begin{figure}
	\begin{center}
		\fig{3.4in}{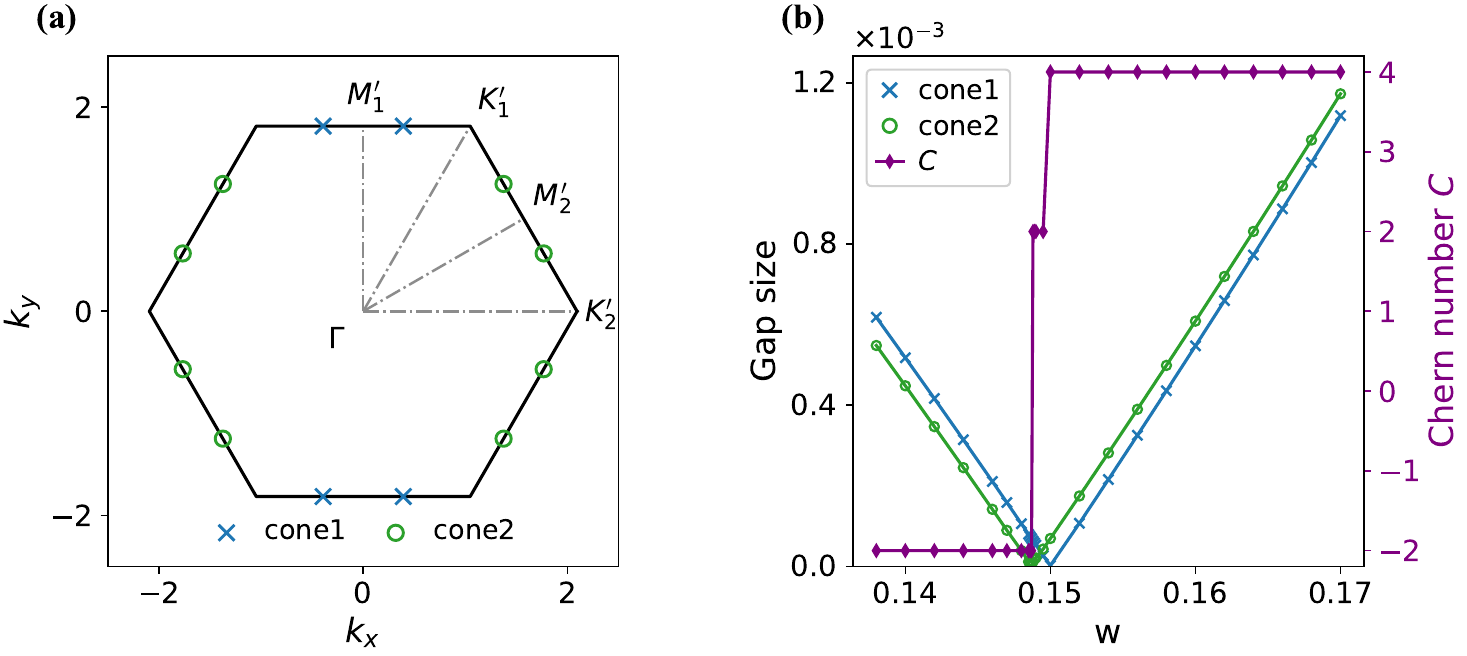} 
		\caption{Gap closings and topological transitions in the cPDW state. (a) Schematic for the two groups of Bogoliubov-Dirac cones in the reduced Brillouin zone. (b) The gap sizes, \textit{i.e.}, local minima of $E_{13,\bk} - E_{12,\bk}$, at the two types of cones as a function of onsite attraction $W$ along the cut at $n=5.7/12$, $J=1.5$ shown in Fig.~\ref{fig2}(b). The purple line displays the corresponding Chern number $C$. } \label{fig5}
	\end{center}
	\vskip-0.8cm
\end{figure}

The small gaps at the Bogoliubov Dirac cones govern the topology of cPDW. 
Along the parameter cut in Fig.~\ref{fig2}(b), increasing $W$ drives two topological transitions while the state remains cPDW. 
Of the six cones in Fig.~\ref{fig5}(a), a group of four closes and reopens its direct gap first, changing the Chern number of the occupied Bogoliubov bands from $C=-2$ to $C=2$ \cite{SZ-PRL08}; the remaining two close and reopen at a slightly larger $W$, changing it to $C=4$ [Fig.~\ref{fig5}(b)]. 
These changes are consistent with one unit of Chern transfer per cone. 
The four-plus-two grouping is consistent with weak rotational-symmetry lowering permitted by the coexistence of the chiral $E_2$ and small $A_1$ components in the self-consistent cPDW state. 
Thus, onsite pairing tunes the gaps of the Bogoliubov Dirac cones and drives successive topological transitions within the cPDW phase.

The tPDW state also breaks time-reversal symmetry, but its three $F_1$ components remain mutually in phase and acquire a common phase difference relative to the uniform $A_1$ condensate, rather than developing the internal phase winding of cPDW. 
This TRS-breaking phase difference produces nonzero Berry curvature on individual Bogoliubov bands at some momenta in our calculation. 
Nevertheless, tPDW retains vertical mirror symmetries, which make the Berry curvature summed over occupied bands opposite at mirror-related momenta and thus enforce $C$ = 0 throughout the gapped tPDW region studied here. 
The $F_1$ phase winding of cPDW breaks these mirrors and permits a nonzero Chern number, although it does not fix its value: $C$ changes across the $W$-driven gap closings within cPDW, as shown in Fig. \ref{fig5}(b)

\subsection{Ginzburg--Landau Analysis}\label{sec:gl}
The numerical phase diagrams suggest that the nPDW and tPDW phases should be understood as new minima created when the uniform and PDW channels become comparable. 
This can be formulated in a Ginzburg--Landau (GL) theory in terms of the uniform $A_1$ order parameter $\psi=\psi^s$ and the three-component bond-$F_1$ PDW order ${\bm\eta}=(\eta_1,\eta_2,\eta_3)={\bm\eta}^b$. We write
\begin{equation}
\mathcal F=\mathcal F_{\rm u}(\psi)+\mathcal F_{\rm PDW}({\bm\eta})+\mathcal F_{\rm int}(\psi,{\bm\eta}),
\end{equation}
where $\mathcal F_{\rm PDW}$ contains the invariants that select the rPDW and cPDW minima in the pure bond-pairing problem \cite{Yao2024PDW, ZhangGM-PRB25}. 
The leading coexistence terms include
\begin{equation}
\begin{split}
\mathcal F_{\rm int}={}&\lambda_1|\psi|^2\sum_\alpha|\eta_\alpha|^2\\
+&\biggl[\lambda_2\psi^{*2}\sum_\alpha\eta_\alpha^2
+\lambda_3\psi^*\sum_{\rm cyc}\eta_\alpha^*\eta_\beta\eta_\gamma
+\mathrm{c.c.}\biggr]+\cdots .
\end{split}
\label{eq:GLcoupling}
\end{equation}
Here $(\alpha,\beta,\gamma)$ runs over cyclic permutations of $(1,2,3)$. The first term controls the overall competition or cooperation between uniform and finite-momentum pairing, while the other two are sensitive to their relative phases. The latter are allowed by the commensurate $M$-point wave vectors: $2\bQ_\alpha$ and $\bQ_\beta+\bQ_\gamma-\bQ_\alpha$ are reciprocal vectors.
The coefficients $\lambda_{1,2,3}$ are phenomenological.
In the following, we show how these symmetry-allowed couplings can favor the observed order-parameter structures for appropriate coefficients, without determining the phase boundaries. 
We note that the onsite amplitude $\psi$ is real and positive, and it is an anomalous average that can be induced at $W$ = 0, even though the onsite pairing term in the BdG Hamiltonian then vanishes.

To interpret nPDW, we consider
\begin{equation}
{\bm\eta}=(a,be^{i\varphi_1},be^{i\varphi_2}),\qquad a,b>0.
\end{equation}
The two phase-sensitive couplings in Eq.~\eqref{eq:GLcoupling} give
\begin{align}
\mathcal F_{\rm phase}={}&2\lambda_2\psi^2
\left[a^2+b^2(\cos2\varphi_1+\cos2\varphi_2)\right]\nonumber\\
+&2\lambda_3\psi ab^2
\left[\cos(\varphi_1+\varphi_2)+2\cos(\varphi_1-\varphi_2)\right].
\label{eq:GLnphase}
\end{align}
For $\lambda_2>0$, the $\lambda_2$ term considered alone favors $\varphi_{1,2}=\pm\pi/2$ and, at fixed $a^2+2b^2$, favors transferring weight from $a$ to $b$. 
Among these phase choices, $\lambda_3>0$ favors $\varphi_1=-\varphi_2$, consistent with the $(a,ib,-ib)$ form of the self-consistent nPDW. 
The intrinsic PDW terms also contribute to the minimum and determine the finite amplitudes and phase boundaries.

The tPDW state instead reflects the phase-sensitive part of the coexistence energy. 
For an equal-amplitude triplet, 
\begin{equation}
{\bm \eta} = \eta e^{i\varphi} (1, 1, 1), \qquad \eta > 0,
\end{equation}
the two phase-sensitive couplings in Eq. (\ref{eq:GLcoupling}) give
\begin{equation}
\mathcal F_{\rm phase} = 6\lambda_2 \psi^2\eta^2 \cos 2\varphi + 6\lambda_3 \psi \eta^3 \cos \varphi.
\end{equation}
For $\lambda_2>0$, the minimum of the above phase-dependent energy satisfies $\cos\varphi=-\lambda_3\eta/(4\lambda_2\psi)$ when $|\lambda_3|\eta<4\lambda_2\psi$; otherwise it occurs at $\varphi=0$ or $\pi$. 
In the former regime, the positive $\lambda_2$ term favors a phase near $\pm\pi/2$, shifted by the $\lambda_3$ term, consistent with the self-consistent tPDW in Table~\ref{tab1}.
Such a state breaks TRS through the relative phase between the PDW and uniform condensates, unlike cPDW, where TRS breaking originates from phase winding within the PDW sector.

The coexistence of $A_1$ and $F_1$ also permits a uniform $E_2$ component. 
Since $2\bQ_\alpha$ is reciprocal, the squares of the three $F_1$ components are translation invariant and contain an $E_2$ combination. 
Multiplication by $\psi^*$ gives this combination the gauge charge of a pair field, allowing it to induce $E_2$ pairing. 
Commensurate cubic PDW combinations provide another allowed source when $\psi$ is very small, as in cPDW \cite{PhysRevResearch.7.023129}. 
Coexistence of chiral $E_2$ with a small $A_1$ component then permits weak rotational-symmetry lowering, consistent with the separation of the Bogoliubov-Dirac cones discussed above.

The GL picture therefore provides a unified interpretation of Fig.~\ref{fig2}. 
At small $W$, $\mathcal F_{\rm PDW}$ dominates and the system remains in one of the parent rPDW or cPDW minima. 
At large $W$, uniform pairing wins and suppresses the finite-\bQ\ order. 
In the intervening regime, the couplings between $\psi$ and ${\bm\eta}$ reshape the PDW free-energy landscape, allowing the nPDW and tPDW minima to appear. 
The central effect of coexistence is thus not a simple superposition of two condensates, but a qualitative reorganization of the internal amplitude and phase structure of the triple-\bQ\ PDW.

\section{Summary and Discussion}
We have studied the interplay between PDW order and uniform superconductivity on the kagome lattice within an effective $t$-$W$-$J$ model containing attractive pairing interactions on NN bonds and lattice sites. 
Starting from the real and chiral triple-$\bQ$ PDW states of the pure bond-pairing limit, we find that increasing the onsite pairing attraction produces an extended coexistence regime before the PDW is eventually suppressed in favor of a translation-invariant $A_1$ superconductor. 
When the finite- and zero-momentum pairing channels become comparable, two additional phases emerge: a nematic PDW state that spontaneously breaks rotational symmetry and a phase-twisted PDW state that breaks TRS through the relative phase between the PDW and uniform condensates.

The coexistence phases are not simple superpositions of the parent orders. Their distinct order-parameter structures generate characteristic charge modulations and quasiparticle spectra.
Within the TRS-breaking cPDW phase, increasing the onsite attraction closes and reopens gaps at the Bogoliubov-Dirac cones, producing successive transitions with Chern number $C=-2$, 2, and 4 along the representative parameter cut. 
The tPDW also breaks TRS, and its Bogoliubov bands have nonzero Berry curvature at some momenta in our calculation; nevertheless, its retained vertical mirror symmetries enfoce $C$ = 0 throughout the gapped tPDW region studied here.
The GL analysis shows that these phenomena can be understood as consequences of symmetry-allowed couplings between the uniform $A_1$ order and the triple-$\bQ$ $F_1$ PDW order: phase-sensitive couplings reorganize both the amplitudes and relative phases of the PDW components. 
Commensurate PDW composites also allow uniform $E_2$ pairing, whose coexistence with $A_1$ can weakly lower rotational symmetry in cPDW. Uniform superconductivity therefore qualitatively reshapes both the symmetry and topology of the PDW manifold.

The kagome vH structure and sublattice interference make this interplay particularly favorable, but the present single-orbital model is intended to isolate its symmetry and pairing mechanisms rather than reproduce the electronic structure of $A$V$_3$Sb$_5$ quantitatively. 
The experiment of Ref.~\cite{Deng2024} reports a U-shaped tunneling $dI/dV$ spectrum with flat residual in-gap states and arc-like quasiparticle-interference signals. 
Its proposed interpretation involves charge-order-reconstructed V $3d$ states, Sb $5p$ states, and orbital-selective uniform and interorbital PDW pairing. 
By comparison, the arc-like zero-energy spectral intensity in our rPDW and cPDW solutions appears with broadening around small but finite quasiparticle gaps, and their DOS is roughly V-shaped outside those gaps. 
These calculated features should therefore not be identified with the observed residual Fermi arcs or the measured tunneling lineshape.

Recent impurity measurements provide another point of contact. 
It has been found that dilute nonmagnetic Ta substitution strongly suppresses the $2\times2$ pairing modulations while uniform superconductivity and charge order remain robust, which is interpreted as indirect phase-sensitive evidence for a sign-modulating PDW \cite{YinJX-PRB26, YinJx-PNAS2026}. 
The finite-$\bQ$ $F_1$ components in our solutions likewise change sign under appropriate lattice translations, and the model allows uniform and PDW orders to coexist while responding differently to changes in pairing strength. 
Our clean calculation does not address impurity pair breaking or identify which coexistence phase occurs in the material \cite{LeeDH-PRB24}. 
A multiorbital treatment incorporating charge-order reconstruction and disorder would be needed to compare the impurity response, tunneling spectrum, and quasiparticle-interference arcs with experiment.

\appendix
\section{Symmetry decomposition of the pairing fields}
\label{app:symmetry}
\subsection{Geometry and extended point group}
A site on sublattice $\alpha=1,2,3$ in the primitive unit cell at $\br=m_1\ba_1+m_2\ba_2$ is located at $\br+\boldsymbol\delta_\alpha$, where
\begin{equation}
	\boldsymbol\delta_1=(0,0),\quad
	\boldsymbol\delta_2=\left(-\tfrac14,-\tfrac{\sqrt3}{4}\right),\quad
	\boldsymbol\delta_3=\left(\tfrac14,-\tfrac{\sqrt3}{4}\right).
\end{equation}
The primitive vectors and $\bQ_\alpha=\bM_\alpha$ are defined in Fig.~\ref{fig1}. For a cyclic permutation $(\alpha,\beta,\gamma)$ of $(1,2,3)$, the bond $(\br,\alpha,+)$ connects sites $(\br,\beta)$ and $(\br,\gamma)$, while $(\br,\alpha,-)$ connects $(\br,\beta)$ and $(\br-\ba_\alpha,\gamma)$. 
Their displacements are $\pm\ba_\alpha/2$. Thus $\nu=\pm1$ labels the two distinct undirected bonds along direction $\ba_\alpha$. Within one $2\times2$ supercell, sums over $\br$ run over $0,\ba_2,\ba_1,\ba_1+\ba_2$.

The extended point group $C'''_{6v}$ classifies zero- and finite-momentum orders on the same footing by retaining primitive translations modulo the supercell translations $2\ba_1$ and $2\ba_2$ \cite{Venderbos2016ExtendedPointGroup, Wagner2023BondFluxKagome, PhysRevResearch.7.023129}. Its four translation classes enlarge the group to order 48. 
In addition to the one-dimensional representations $A_1,A_2,B_1,B_2$ and two-dimensional representations $E_1,E_2$ at zero momentum, it contains four three-dimensional representations $F_1,\ldots,F_4$ associated with finite-$\bQ$ modulations. 
We define
\begin{equation}
	s_\kappa(\br)=\cos(\bQ_\kappa\cdot\br),\qquad \kappa=1,2,3.
\end{equation}
The argument contains the primitive-cell position $\br$, without the sublattice displacement. 
The resulting signs within one supercell are
\begin{equation}
	\begin{array}{c|rrr}
		\br&s_1&s_2&s_3\\\hline
		0&1&1&1\\
		\ba_2&-1&1&-1\\
		\ba_1&1&-1&-1\\
		\ba_1+\ba_2&-1&-1&1
	\end{array}
	\label{eq:signs}
\end{equation}
and fix the origin convention used throughout the manuscript.

\subsection{Symmetry projection of the pairing fields}
Let $p\in\{s,b\}$ specify the pairing sector, with $s$ denoting onsite pairing and $b$ denoting nearest-neighbor bond pairing. 
A position label $x$ refers to a site when $p=s$ and to an undirected bond when $p=b$. More explicitly, the sets of labels in one supercell are
\begin{align}
	\mathcal X_s&={(\br,\alpha):\br\in\mathcal R,\ \alpha=1,2,3},
	&N_s&=|\mathcal X_s|=12, \nonumber\\
	\mathcal X_b&={(\br,\alpha,\nu):\br\in\mathcal R,
		\alpha=1,2,3,\ \nu=\pm1},
	&N_b&=|\mathcal X_b|=24,
\end{align}
where $\mathcal R={0,\ba_2,\ba_1,\ba_1+\ba_2}$. 
Accordingly, the two pairing fields are vectors $\boldsymbol\Delta^s\in\mathbb C^{12}$ and $\boldsymbol\Delta^b\in\mathbb C^{24}$, with components $\Delta^s(\br,\alpha) = \Delta^s_{\br,\alpha}$ and $\Delta^b(\br,\alpha,\nu) =\Delta^b_{\br,\alpha,\nu}$.

For either sector, the inner product of two patterns is
\begin{equation}
	\langle u|v\rangle_p
	=\sum_{x\in\mathcal X_p}u^*(x)v(x).
\end{equation}
We use real basis patterns $f_\mu^p$, where the abstract label $\mu$ identifies a basis pattern rather than a site, bond, or pairing sector. 
They are normalized such that
\begin{equation}
	\langle f_\mu^p|f_{\mu'}^p\rangle_p
	=N_p\delta_{\mu\mu'}.
\end{equation}
The uniform $A_1$ pattern is $f_{A_1}^p(x)=1$. Its coefficient is denoted by $\psi^p$.

For the two $E_2$ components, labeled by $\ell=1,2$, we use the directional vectors
\begin{equation}
	{\bm e}_1=\tfrac{1}{\sqrt2}(-1,-1,2),\qquad
	{\bm e}_2=\sqrt{\tfrac32}(-1,1,0).
	\label{eq:E2basis}
\end{equation}
They satisfy $\sum_\alpha e_{\ell,\alpha} e_{\ell',\alpha} =3\delta_{\ell\ell'}$. 
The same directional convention is used for sites and bonds:
\begin{align}
	f^s_{E_2,\ell}(\br,\alpha)&=e_{\ell,\alpha},\nonumber\\
	f^b_{E_2,\ell}(\br,\alpha,\nu)&=e_{\ell,\alpha}.
\end{align}
These patterns repeat in every primitive unit cell and therefore have zero momentum.

The selected $F_1$ triplet, labeled by $\kappa=1,2,3$, is
\begin{align}
	f^s_{F_1,\kappa}(\br,\alpha)
	&=\sqrt3\,\delta_{\alpha\kappa}s_\kappa(\br),\nonumber\\
	f^b_{F_1,\kappa}(\br,\alpha,\nu)
	&=\sqrt3\,\delta_{\alpha\kappa}\nu s_\kappa(\br).
	\label{eq:F1basis}
\end{align}
Its components have disjoint support on the three sublattices or bond directions and transform under primitive translations according to Eq.~\eqref{eq:signs}. 
The bond label $\nu$ distinguishes two undirected bonds; it does not specify an ordering of their endpoints. 
The factor $\nu$ is therefore compatible with singlet pairing, $\Delta_{ij}=\Delta_{ji}$. 
All six retained patterns in each sector obey the normalization given above.

The coefficient of any retained pattern is obtained by orthogonal projection,
\begin{equation}
	c_\mu^p
	=\frac{\langle f_\mu^p|\boldsymbol\Delta^p\rangle_p}{N_p}
	=\frac1{N_p}\sum_{x\in\mathcal X_p}
	[f_\mu^p(x)]^*\Delta^p(x).
\end{equation}
Thus $c_{A_1}^p=\psi^p$, $c_{E_2,\ell}^p=\xi_\ell^p$, and $c_{F_1,\kappa}^p=\eta_\kappa^p$. In particular,
\begin{align}
	\psi^s&=\frac1{12}\sum_{\br,\alpha}\Delta^s_{\br,\alpha},
	&
	\psi^b&=\frac1{24}\sum_{\br,\alpha,\nu}
	\Delta^b_{\br,\alpha,\nu},\nonumber\\
	\xi_\ell^s&=\frac1{12}\sum_{\br,\alpha}
	e_{\ell,\alpha}\Delta^s_{\br,\alpha},
	&
	\xi_\ell^b&=\frac1{24}\sum_{\br,\alpha,\nu}
	e_{\ell,\alpha}\Delta^b_{\br,\alpha,\nu}.
\end{align}
For the finite-momentum components, these projections reduce to
\begin{align}
	\eta^s_\kappa
	&=\frac1{4\sqrt3}\sum_{\br}
	s_\kappa(\br)\Delta^s_{\br,\kappa},\nonumber\\
	\eta^b_\kappa
	&=\frac1{8\sqrt3}\sum_{\br,\nu}
	\nu s_\kappa(\br)\Delta^b_{\br,\kappa,\nu}.
	\label{eq:PDWprojection}
\end{align}
All sums over $\br$ in these equations are restricted to $\mathcal R$.

The projected components and the residual field give the exact decompositions
\begin{align}
	\Delta^s_{\br,\alpha}
	={}&\psi^s+\sum_{\ell=1}^2\xi^s_\ell e_{\ell,\alpha}
	+\sqrt3\,\eta^s_\alpha s_\alpha(\br)
	+\Delta^s_{\perp;\br,\alpha},
	\label{app1}\\
	\Delta^b_{\br,\alpha,\nu}
	={}&\psi^b+\sum_{\ell=1}^2\xi^b_\ell e_{\ell,\alpha}
	+\sqrt3\,\nu\eta^b_\alpha s_\alpha(\br)
	+\Delta^b_{\perp;\br,\alpha,\nu}.
	\label{app2}
\end{align}
Here $\boldsymbol\Delta_\perp^p$ is orthogonal to the retained $A_1$, $E_2$, and selected $F_1$ patterns. It contains all other components of the self-consistent field. 
The symmetry projection is used to characterize the solutions; energies, spectra, and topology are calculated from the full fields.

For a nonzero pairing field, the squared-norm weights of the retained sectors and the residual fraction are
\begin{align}
	w_{A_1}^p
	&=\frac{N_p|\psi^p|^2}{|\boldsymbol\Delta^p|^2},
	&
	w_{E_2}^p
	&=\frac{N_p\sum_{\ell=1}^2|\xi_\ell^p|^2}
	{|\boldsymbol\Delta^p|^2},\nonumber\\
	w_{F_1}^p
	&=\frac{N_p\sum_{\kappa=1}^3|\eta_\kappa^p|^2}
	{|\boldsymbol\Delta^p|^2},
	&
	r^p
	&=1-w_{A_1}^p-w_{E_2}^p-w_{F_1}^p.
	\label{eq:weights}
\end{align}
The $F_1$ weight here refers only to the selected triplet in Eq.~\eqref{eq:F1basis}. Orthogonality gives
\begin{equation}
	\frac{|\boldsymbol\Delta^p|^2}{N_p}
	=|\psi^p|^2+\sum_{\ell=1}^2|\xi_\ell^p|^2
	+\sum_{\kappa=1}^3|\eta_\kappa^p|^2
	+\frac{|\boldsymbol\Delta_\perp^p|^2}{N_p},
	\label{eq:parseval}
\end{equation}
so $r^p =|\boldsymbol\Delta_\perp^p|^2/ |\boldsymbol\Delta^p|^2$. 
These quantities measure fractions of the squared norm, rather than pairing amplitude or energy.

The bond space contains a second, orthogonal copy of $F_1$ \cite{PhysRevResearch.7.023129}. 
It can be constructed from the endpoint values of the onsite $F_1$ patterns:
\begin{align}
	\widetilde f^b_{F_1,\kappa}(ij)
	&=\frac{f^s_{F_1,\kappa}(i)+f^s_{F_1,\kappa}(j)}
	{\sqrt2},\nonumber\\
	\widetilde\eta^b_\kappa
	&=\frac1{24}\sum_{\langle ij\rangle\in\mathrm{SC}}
	[\widetilde f^b_{F_1,\kappa}(ij)]^*\Delta_{ij}.
	\label{eq:secondF1}
\end{align}
Each $\widetilde f^b_{F_1,\kappa}$ has squared norm $N_b=24$ and is orthogonal to the selected triplet in Eq.~\eqref{eq:F1basis}. 
Its contribution is included in $\boldsymbol\Delta_\perp^b$, not in $\boldsymbol\eta^b$ or $w_{F_1}^b$.

Finally, all projection coefficients may be complex. We fix their common global $U(1)$ phase by making $\psi^s$ real and positive; if $\psi^s$ vanishes, another nonzero coefficient can fix this phase. 
The onsite and bond sectors must undergo the same gauge rotation to preserve their physical relative phases. 
A basis rotation within a degenerate representation changes its individual coefficients but leaves the reconstructed field and its total weight unchanged. 
We therefore keep the conventions of Eqs.~\eqref{eq:signs}, \eqref{eq:E2basis}, and \eqref{eq:F1basis} fixed when comparing different phases.

\bibliography{referencespaper} 	
\end{document}